\documentclass{article}

\usepackage{PRIMEarxiv}

\usepackage{amsmath}
\usepackage{graphicx}%
\usepackage{multirow}%
\usepackage{amsmath,amssymb,amsfonts}%
\usepackage{amsthm}%
\usepackage{amsthm}%
\usepackage{mathrsfs}%
\usepackage[title]{appendix}%
\usepackage{manyfoot}%
\usepackage{booktabs}%
\usepackage{algorithm}%
\usepackage{algorithmicx}%
\usepackage{algpseudocode}%
\usepackage{listings}%

\title{Tiling Spaces and the Expanding Universe: Bridging Quantum Mechanics and Cosmology}

\author{Marcelo Amaral\textsuperscript{1} and Aman Yadav \textsuperscript{2}\\
  \textsuperscript{1}\emph{Quantum Gravity Research, Los Angeles, CA 90290, USA}\\
\textsuperscript{2}\emph{Delhi
Technological University, Delhi 110042, India}\\
}

\begin{document}
\maketitle

\begin{abstract}
We propose a heuristic model of the universe as a growing quasicrystal projected from a higher-dimensional lattice. This quasicrystalline framework offers a novel perspective on cosmic expansion, where the intrinsic growth dynamics naturally give rise to the observed large-scale expansion of the universe. Motivated by this model, we explore the Schr{\"o}dinger equation for a particle in a box with time-dependent boundaries, representing the expanding underlying space. By introducing a constraint that links microscale quantum phenomena with macroscale cosmological quantities, we derive an equation resembling the Friedmann equation, providing potential insights into the Hubble tension.
Our model incorporates phonons and phasons—quasiparticles inherent in quasicrystalline structures—that play critical roles in cosmic-scale dynamics and the universe’s expansion. This framework suggests that the necessity for an inflationary period may be obviated.
Furthermore, phonons arising from the quasicrystalline structure may serve as dark matter candidates, influencing the dynamics of ordinary matter while remaining largely undetectable through electromagnetic interactions. Drawing parallels with crystalline matter at atomic scales, which is fundamentally quantum in nature, we explore how the notion of tiling space can support continuous symmetry atop a discrete structure. This provides a novel framework for understanding the universe’s expansion and underlying structure.
Consequently, our approach suggests that further development could enhance our understanding of cosmic expansion and the universe's structure, bridging concepts from quantum mechanics, condensed matter physics, and cosmology.
\end{abstract}

\keywords{Quantum Gravity, Cosmology, Hubble Tension, Quasicrystals}

\tableofcontents

\section{Introduction}
\label{intro}

Recent developments in theoretical physics have propelled us toward a deeper understanding of spacetime and gravity. The field of emergent gravity \cite{Verlinde2011} suggests that spacetime and gravity may be byproducts of fundamental quantum mechanical processes. This paradigm shift is supported by theories such as loop quantum gravity and spin foam \cite{Rovelli-Vidotto-Book}, as well as superstring theory and M theory \cite{stringbook1}, all of which propose a quantum-level discreteness underlying gravity.

A key focus since the advent of perturbative superstring theory has been the quest to unravel quantum gravity \cite{Kiefer}. This pursuit requires a deep understanding of the theory’s fundamental principles and extracting insights about Planck-scale physics \cite{Rovelli1998}. Consequently, there is a growing trend to incorporate concepts from condensed matter or solid-state physics, such as viewing spacetime as a deformable solid or crystal \cite{Tahim2009}. More recently, quasicrystals have been investigated \cite{Amaral:2023mgf,modelsetemergence}.

Clues that our theories about reality are incomplete are emerging in various fields, from anomalies in particle physics experiments to unresolved problems in cosmology \cite{lawsofnature}. For instance, the discovery of neutrino oscillations challenged the Standard Model of particle physics, indicating that neutrinos have mass. Similarly, in cosmology, the Hubble tension \cite{hubbletension} represents a profound discrepancy between measurements of the universe's expansion rate. On one side, observations of the Cosmic Microwave Background (CMB) by missions like the European Space Agency's Planck satellite suggest a slower expansion rate, based on the state of the universe shortly after the Big Bang and theoretical models of its evolution—around 67-68 km/s/Mpc. On the other, direct measurements of the current universe, notably from the Hubble Space Telescope and confirmed by the James Webb Space Telescope, indicate a significantly faster rate \cite{hubble}—around 73-74 km/s/Mpc. This tension challenges our understanding of the cosmos and suggests that our theoretical models of the universe may be incomplete, indicating that new physics might be necessary to resolve the discrepancy. Despite the increased precision of observational instruments and techniques, this persistent tension highlights a crucial gap in our understanding of fundamental cosmic processes, making it a focal point for theoretical and observational cosmology alike.

Our research adopts a dual approach within this context. First, we draw inspiration from the concept of a deformable solid or crystal and quantum cosmology, which deals with discreteness, and we propose modeling spacetime as a growing quasicrystal. Characterized by long-range order without periodicity, this structure reflects the aperiodic patterns found in condensed matter physics. Our principal result suggests that the cosmological scale factor can be interpreted as an intrinsic scaling property within a quasicrystal tiling space framework. We further explore the implications of this model for resolving the Hubble tension, offering a novel perspective on the universe's evolution and providing insights into other unresolved issues in cosmological models, including dark matter and dark energy. A critical aspect of quasicrystal growth is that external expansion may be accompanied by internal subdivision and rescaling, which parallels the observed expansion of the universe, where space appears to be generated between galaxies.

Second, we build upon conventional and relativistic quantum mechanics. While the universe's expansion is traditionally addressed from a classical perspective on large scales, this expansion must also extend down to microscopic scales, as suggested by various approaches in quantum cosmology \cite{bojowald,Kuzmichev} and by our quasicrystal growth model. Although Hubble-Lemaître's law is empirically validated at cosmological distances, we extend its conceptual framework to `intermediate' and even microscopic scales as a theoretical proposition. While these scales are not typically associated with cosmic expansion, our quasicrystal model of spacetime provides a basis for such an extension. This hypothesis enables us to explore the implications of spacetime expansion on quantum states confined within matter-dominated regions.

We analyze the wave function of a particle with mass $m$ within a specific spatial domain using both the Schr{\"o}dinger equation and the Klein-Gordon equation. Crucially, we posit that the wavenumber, wavelength, and quantum amplitudes become functions of the dynamic size of the spatial boundary and are explicitly subject to the Hubble-Lemaître's law. Considerations of the evolving boundary of this spatial domain lead us to a modified version of the Friedmann equation, a key component in cosmological models, thereby highlighting the emergence of classical spacetime from quantum mechanical principles. We apply this framework to the later stages of the universe's evolution under conventional quantum mechanics to address the global growth and rescaling of the quasicrystal structure. The early universe, where curved spacetime geometry plays a significant role, is left as a subject for future work.

Our research converges on the concept of rescaling the underlying geometry along with emerging constraints relevant to physics at different scales. This approach aligns with our framework of conceptualizing spacetime as a quasicrystalline structure. By incorporating scale invariance, we aim to establish a connection between the microscale domains of quantum mechanics and the macroscale phenomena observed in cosmology. This synthesis provides a unified perspective that may offer deeper insights into the fundamental nature of the universe.

In the subsequent sections, we will explore the notion of spacetime as a tiling space, drawing parallels with quasiperiodic structures in condensed matter physics to strengthen our hypothesis regarding emergent spacetime and gravity. Additionally, we will expand on the mathematical foundations of our model, examining how energy considerations translate into cosmological parameters.

\section{The Universe Expansion from Quasicrystalline Tiling Spaces}
\label{sec:freedmaneqqc}
Quasicrystals are characterized by several distinctive properties that make them a compelling model for the universe. These properties, as discussed in \cite{BaakeGrimm}, include: 

\begin{itemize}
    \item \textbf{Uniformly Discrete:} For a certain radius \( r > 0 \), every sphere of radius \( r \) contains at most one point of the quasicrystal point set, \( \triangle \).
    \item \textbf{Relatively Dense:} There exists a radius \( R > 0 \) such that each sphere of this radius encompasses at least one point of \( \triangle \).
    \item \textbf{Locally Finite:} For any compact subset \( \varLambda \) in \( \mathbb{R}^{d} \), the intersection \( C = \varLambda \cap \triangle \), or cluster, is either finite or empty.
    \item \textbf{Finite Local Complexity (FLC):} The set \( \{ (t + \varLambda) \cap \triangle | t \in \mathbb{R}^{d} \} \), often represented by translations \( \triangle - \triangle \), comprises only a finite number of clusters, modulo translations.
    \item \textbf{Uniform Distribution:} As proven in \cite{BaakeGrimm}, the distribution of points within quasicrystals is uniform.
\end{itemize}
The first four properties align with a universe where matter is neither infinitely dense nor arbitrarily dispersed, and where the arrangement of matter is limited to a finite number of configurations. The fifth property, uniform distribution, supports the cosmological principle by suggesting large-scale homogeneity and isotropy in the universe, while allowing for small inhomogeneities.

A key aspect of quasicrystals with the aforementioned properties is their derivation from a specific projection from a higher-dimensional lattice. The so-called cut-and-project scheme (CPS) is defined as a 3-tuple \( \mathcal{G} = (\mathbb{R}^{d}, G, \mathcal{L}) \), where \( \mathbb{R}^{d} \) represents a real Euclidean space, \( G \) is a locally compact abelian group (or any topological group), and \( \mathcal{L} \) is a lattice in \( \mathbb{R}^{d} \times G \). For our purposes, this scheme includes a natural projection \( \pi: \mathbb{R}^{d} \times G \rightarrow \mathbb{R}^{d} \). In the context of an infinite quasicrystal, the projection typically involves restricting the points to a non-empty, relatively compact subset \( K \subset G \), referred to as the window. However, in our study of a finite universe and finite quasicrystals, we project the entire mother lattice onto \( \mathbb{R}^{d} \) part by part as explained below, thereby obviating the need to consider this window.

To illustrate our approach more concretely, let us consider the 4D Elser-Sloane quasicrystal (ESQC) derived from the E8 root lattice \cite{ElserSloane1987}, which is the most studied 4D quasicrystal. We position a coordinate system on the E8 lattice, with the origin placed at an E8 vertex. The E8 lattice can then be visualized as concentric shells of points radiating from the origin. As noted in \cite{ConwaySloane1988}, the number of points on each E8 shell increases in a nonlinear fashion, described by the function 
\begin{equation}
N_{E8}(n)=240\sum_{d|n}d^{3}
\end{equation}
where the right-hand side of the equation is the sum of the cubes of all divisors $d$ of $n$, multiplied by 240. Projecting shell by shell, the E8 lattice to 4D using the ESQC CPS generates a growing 4D quasicrystal.

\begin{figure}[ht]
    \centering
    \includegraphics[scale=0.40]{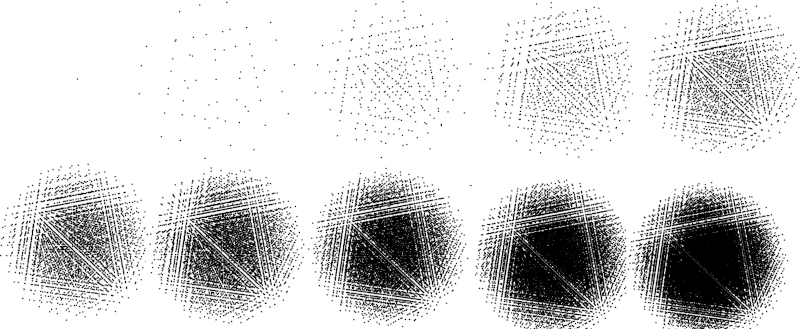}
    \caption{First 10 shells from E8 lattice to 4D ESQC to a 3D slice.}
    \label{e8to3dfirstshells}
\end{figure}

The projection of the initial shells of the E8 lattice, when viewed from a 3D slice perspective, reveals distinct patterns that emerge through rescaling. In this process, the quasicrystal grows externally in size while simultaneously undergoing internal subdivision. See Figure (\ref{e8to3dfirstshells}) for a visual representation.

The notion of tiling spaces emerges by treating each possible tiling configuration that can be constructed on a specific CPS as a point in this space. At each stage of the growth process, different tiling configurations are possible. Quantum mechanically, each possible tiling can be represented as a state, allowing us to describe the growth using the formalism of Hilbert spaces \cite{Connes}. As we will show, tiling spaces offer a robust framework for capturing both the geometric and quantum mechanical properties of the universe's expansion.

\subsection{Universe as a Growing Quasicrystal Hypothesis}

Our working hypothesis is that the universe is a growing quasicrystal. This is a simple heuristic model of the universe. Let us gather support for this idea and explore its consequences.

\begin{figure}[ht]
    \centering
    \includegraphics[scale=0.40]{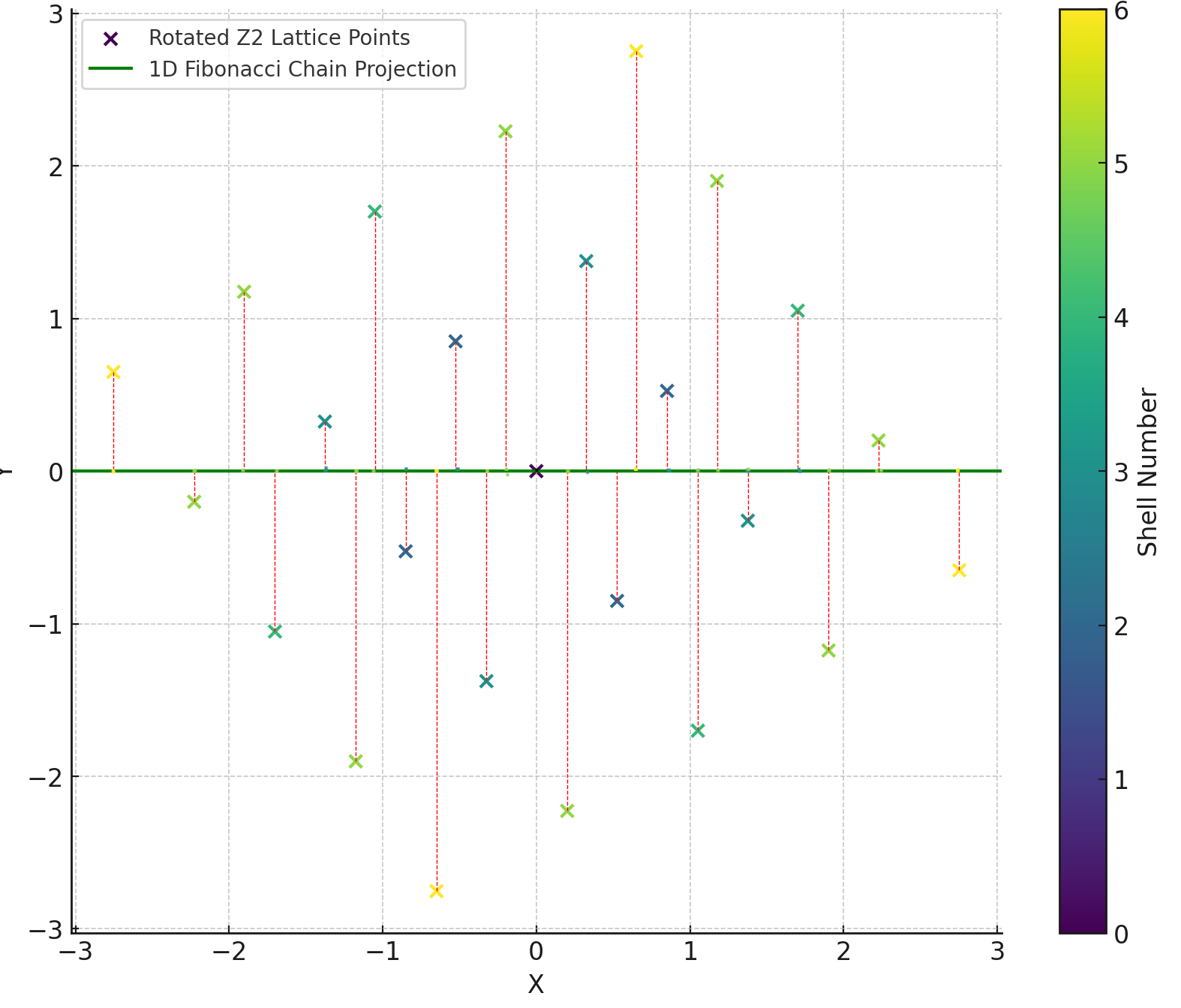}
    \caption{A rotated Z2 is projected shell by shell with the color tracking the ordering. As this one‐dimensional universe expands, it also gets new points within the already projected region.}
    \label{fibchain}
\end{figure}

This model proposes an unconventional origin of the universe, diverging from the traditional Big Bang theory. It suggests that the universe began with the projection of a single point from a higher-dimensional lattice, such as the E8 root lattice. This initiated a sequence of point projections that sequentially constructed the universe's structure. This process, governed by a computational framework, unfolds step by step transitioning from one shell $N_{\text{E8}}(n)$ to the next $N_{\text{E8}}(n+1)$ and so forth. The number of points in each shell grows rapidly, with the first ten shells containing {240, 2160, 6720, 17520, 30240, 60480, 82560, 140400, 181680, 272160} points, respectively.  
The projection into a 4D quasicrystalline structure not only expands the universe externally but also subdivides it internally. Points from newer shells intermingle with those from older shells, illustrating a complex process of growth and subdivision characteristic of the early universe (see Figure (\ref{fibchain}) for a lower-dimensional illustration and Figure (\ref{e8to3dfirstshells}) for 3D slices of the 4D ESQC).

In this early stage, the universe consists of points, which are projections from the higher-dimensional root lattice. In the Lie algebra framework of modern gauge theories, these root lattice points represent fundamental particle states that form the Hilbert spaces of representations of the Lie algebra. At this stage, there is no scale or mass, and thus no distinction between particles, highlighting an analogy to the grand unification scale.

In condensed matter physics, a growing quasicrystal is associated with expanding Hilbert spaces. We can conjecture a similar process for the universe. As the universe grows, a burgeoning Hilbert space $\mathcal{H}_n$ emerges at each growth step $n$, reflecting the tiling spaces \cite{sadun2018} of potential quasicrystalline configurations. To address the different tiling configurations of quasicrystals, it is standard to consider the associated \( C^* \)-algebra structures \cite{Connes} and the notion of tiling spaces. This Hilbert space can be constructed as follows: Let $\mathcal{T}_n$ denote the set of all distinct tiling configurations at the $n$-th growth step, with each tiling configuration $T_n \in \mathcal{T}_n$ represented as a basis state $|T_n\rangle$ in $\mathcal{H}_n$. The growth process from step $n$ to $n+1$ can be expressed as:
\begin{align}
\mathcal{T}_n &\rightarrow \mathcal{T}_{n+1}, \\
\mathcal{H}_n &\rightarrow \mathcal{H}_{n+1}.
\end{align}

The wavefunction of the universe at the $n$-th growth step, $|\Psi_n\rangle \in \mathcal{H}_n$, can be written as a superposition of all possible tiling configurations:
\begin{equation}
|\Psi_n\rangle = \sum_{T_n \in \mathcal{T}_n} \alpha(T_n) |T_n\rangle,
\end{equation}
where $\alpha(T_n)$ are complex coefficients determined by the specific growth rules of the quasicrystal. As the universe grows, the wavefunction evolves to incorporate the new tiling configurations that emerge at each step, $|\Psi_n\rangle \rightarrow |\Psi_{n+1}\rangle$.

Crystals and quasicrystals in condensed matter physics or solid state systems are known for their phonon quasiparticles, which emerge due to the breaking of translational symmetry. Quasicrystals, however, also host phasons—a less understood type of quasiparticle. 

As the quasicrystalline universe evolves, phonons and phasons should emerge naturally due to the intrinsic properties of growing quasicrystals, even those composed of 'spacetime atoms.' Phonons correspond to collective vibrational modes of the tiling configurations, while phasons are associated with the rearrangement of the quasicrystalline structure without changing the overall energy. The presence of these quasiparticles is significant, as their dispersion relations lead to a form of the Friedmann equation Equation (\ref{phasonH2}), implying the establishment of standard expansion.

To incorporate phonons and phasons into our quantum description of tiling spaces, we can introduce creation and annihilation operators acting on the Hilbert space $\mathcal{H}_n$ at each growth step. Let $a^{\dagger}_{k,n}$ and $a_{k,n}$ denote the creation and annihilation operators for a phonon with wave vector $k$ at the $n$-th growth step, satisfying the commutation relations:
\begin{equation}
[a_{k,n}, a^{\dagger}_{k',n}] = \delta_{k,k'}, \quad [a_{k,n}, a_{k',n}] = [a^{\dagger}_{k,n}, a^{\dagger}_{k',n}] = 0.
\end{equation}

Similarly, let $b^{\dagger}_{q,n}$ and $b_{q,n}$ denote the creation and annihilation operators for a phason with wave vector $q$ at the $n$-th growth step, satisfying similar commutation relations:
\begin{equation}
[b_{q,n}, b^{\dagger}_{q',n}] = \delta_{q,q'}, \quad [b_{q,n}, b_{q',n}] = [b^{\dagger}_{q,n}, b^{\dagger}_{q',n}] = 0.
\end{equation}
The phonon and phason operators commute with each other:
\begin{equation}
[a_{k,n}, b_{q,n}] = [a_{k,n}, b^{\dagger}_{q,n}] = [a^{\dagger}_{k,n}, b_{q,n}] = [a^{\dagger}_{k,n}, b^{\dagger}_{q,n}] = 0.
\end{equation}
Using these operators, we define phonon and phason excited states by acting on the tiling configuration states $|T_n\rangle$:
\begin{align}
|T_n; \{k\}_m; \{q\}_l\rangle &= a^{\dagger}_{k_1,n} \ldots a^{\dagger}_{k_m,n} b^{\dagger}_{q_1,n} \ldots b^{\dagger}_{q_l,n} |T_n\rangle,
\end{align}
where $\{k\}_m = \{k_1, \ldots, k_m\}$ and $\{q\}_l = \{q_1, \ldots, q_l\}$ characterize the $m$ phonons and $l$ phasons, respectively.

The growth of the quasicrystalline universe can be described in terms of the evolution of these excited states. As the universe grows from step $n$ to step $n+1$, the phonon and phason creation and annihilation operators evolve:
\begin{align}
a^{\dagger}_{k,n} &\rightarrow a^{\dagger}_{k,n+1}, \quad a_{k,n} \rightarrow a_{k,n+1}, \\
b^{\dagger}_{q,n} &\rightarrow b^{\dagger}_{q,n+1}, \quad b_{q,n} \rightarrow b_{q,n+1}.
\end{align}
The wavefunctions $|\Psi_n\rangle$ and $|\Psi_{n,[T]}\rangle$ can be expanded in terms of these excited states:
\begin{align}
|\Psi_n\rangle &= \sum_{T_n \in \mathcal{T}_n} \sum_{\{k\}, \{q\}} \alpha(T_n; \{k\}, \{q\}) |T_n; \{k\}; \{q\}\rangle, \\
|\Psi_{n,[T]}\rangle &= \sum_{g \in G_n} \sum_{\{k\}, \{q\}} \beta(g T_n; \{k\}, \{q\}) |g T_n; \{k\}; \{q\}\rangle,
\end{align}
where the coefficients $\alpha(T_n; \{k\}, \{q\})$ and $\beta(g T_n; \{k\}, \{q\})$ depend on both the tiling configurations and the phonon and phason excitations.

\subsection{Tiling Space}

Let us switch now to a slight different formalist to demonstrate in a straightforward manner how energy density arises naturally from the underlying discrete structure in the regime of large number of points and tiles, which should be the regime where phonons and phasons emerge. We can start with the notion of tiling spaces using the probabilistic approach outlined in \cite{BaakeGrimm}. Tiling spaces are defined by the continuous hull $X(\varLambda)$, which includes all translations of a point set $\varLambda$ in $\mathbb{R}^{d}$. This framework is typically applied to infinite quasicrystals. However, to understand regimes where dynamics occur at scales much larger than the underlying spacing (potentially at the Planck scale), we expect to recover invariance under continuous translations in the tiling space.

A characteristic property of a point set in $\mathbb{R}^{d}$ is the average number of points per unit volume, and we are interested in the relative frequency $f^{r}$ of subsets or clusters of points, defined as the absolute frequency per unit volume over the density of the point set $\varLambda$. Various methods to compute this quantity can be found in the quasicrystal literature \cite{BaakeGrimm}. This will allow us to define a probabilistic measure in the tiling space.

These relative frequencies define an invariant probability measure on the set of tiling configurations $X_{0}(\varLambda)$, which consists of all tiling configurations with the origin as a point. For a non-empty finite cluster $P$, we define the cylinder set $Z_{P}$ containing all configurations with $P$ as a subcluster:
\begin{equation}
Z_{P}:=\{\varLambda'\in X_{0}(\varLambda)\mid P\subset\varLambda'\}.
\end{equation}
The measure of this cylinder set is given by the relative frequency:
\begin{equation}
\mu_{0}(Z_{P})=f_{\varLambda}^{r}(P).
\end{equation}

This measure extends to a positive measure on the entire tiling space $X(\varLambda)$ by standard arguments, using the \(\sigma\)-algebra generated by the cylinder sets.  
The invariant measure$\mu$ provides a robust framework to explore the statistical properties of quasicrystalline structures and their implications for cosmology. We can express the partition function as: 
\begin{align}
Z_{X} & =\sum_{\varLambda\in X(\varLambda)}\prod_{P\in\varLambda}f_{\varLambda}^{r}(P)=\sum_{\varLambda\in X(\varLambda)}\prod_{P\in\varLambda}\left(f_{\varLambda}^{r}\right)^{Nf_{\varLambda}^{r}}\nonumber \\
 & =\sum_{\varLambda\in X(\varLambda)}e^{N\sum_{P}f_{\varLambda}^{r}(P)\ln(f_{\varLambda}^{r}(P))}=\sum_{\varLambda\in X(\varLambda)}e^{-NI_{\varLambda}},
\end{align}
where $N$ is large, implying that the number of specific configurations
$P$ or clusters must approach its frequency. We introduce the information
entropy of the tiling space as 
\begin{equation}
I_{\varLambda}=-\sum_{P}f_{\varLambda}^{r}(P)\ln(f_{\varLambda}^{r}(P)).
\end{equation}

To connect this framework with a notion of energy density, we start
by calculating the emergent average energy $\langle E\rangle$ from
the partition function $Z$: 
\begin{equation}
\langle E\rangle=-\frac{\partial lnZ_{X}}{\partial\beta},
\end{equation}
where $\beta=N=\frac{1}{kT}$ is the inverse temperature. Note that
$N$ is playing the role of $\beta$. As the tiling space grows the
temperature lowers.

Given the volume $\eta$, the energy density $\rho$ is then: 
\begin{equation}
\rho=\frac{\langle E\rangle}{\eta}.
\end{equation}

This framework allows us to analyze the statistical properties of quasicrystalline structures and their implications for cosmology. The information entropy $I_{\varLambda}$ plays a crucial role in understanding the distribution and frequency of clusters within the tiling space, and the derived energy density connects this microscopic structure to macroscopic physical properties at large scales. 

\subsection{Phonons , Phasons, Effective Field Theory}

Thus, the early growing quasicrystal universe can be described microscopically using quantum mechanics or the distribution and frequency of clusters within the tiling spaces as outlined above. 
However, when the universe becomes large enough to support phonons and phasons, the regime of large number of tiles mentioned above, a new phase emerge. 
When phonons and phasons are present, knowledge from solid state physics tell us that in the regime of short wavelengths, the dispersion relation becomes nonlinear, and the microscopic details of the lattice become important. Phonons become sensitive to the discrete nature of the lattice, and the nonlinear dispersion can lead to phonon dispersion, where phonon packets with different wavelengths propagate at different speeds, causing distortion of the waveform. 
However, the dispersion relation for phonons in lattices is similar to the relativistic dispersion relation in the regime of large wavelengths, but with the speed of light replaced by the speed of sound. In this regime, the phonon dispersion relation is nearly linear, and the crystal can be treated as a continuous elastic medium, with phonons behaving like classical sound waves. 
The transition between these two regimes occurs when the phonon wavelength becomes comparable to the interatomic spacing, with the specific wavevector or wavelength at which this transition happens depending on the details of the crystal structure and the interatomic forces.

In quasicrystals, the situation is more complex because they do not have a simple primitive cell like periodic crystals but rather a complex, self-similar structure. As a result, the phonon dispersion relations in quasicrystals can exhibit intricate, fractal-like patterns. In the context of cosmology, self-similar structures are reminiscent of fractal patterns, and there has been significant interest in exploring fractal approaches to general relativity and cosmology \cite{Barbour2003, Pietronero1987, Labini1998, Joyce2005}. These studies investigate the possibility that the distribution of matter in the universe exhibits fractal properties over a range of scales. By connecting our quasicrystal model to fractal cosmology, we align with previous efforts to describe the universe's large-scale structure using self-similar patterns.

Despite the differences between phonons in crystals and quasicrystals, the linear dispersion relation for phonons in quasicrystals shares similarities with the phonon branches in periodic crystals, particularly at low frequencies and long wavelengths. Phasons in quasicrystals also share similarities with phenomena found in relativistic physics. However, a key difference is the presence of a dispersive imaginary term in the phason dispersion relation, which becomes prominent at very large wavelengths \cite{Baggioli:2020haa, Lubensky}. This imaginary term is related to the diffusive behavior of phasons and is absent in the phonon dispersion relation.

We can extend our discussion by proposing an effective field theory (EFT) for quasicrystals that provides a concrete mechanism for setting the inflaton field, offering an alternative to standard reheating mechanisms. Following the approach in \cite{Baggioli:2020haa}, we consider the phason field inherent in quasicrystals as a candidate for the inflaton.

In the regime of large wavelengths, the quasicrystal can be treated as a continuous elastic medium. This approximation allows us to model the quasicrystal using scalar fields within an EFT framework, aiming to reproduce the results of conventional inflationary models. As quasicrystals exist at finite temperatures, their equilibrium state is described by a thermal density matrix involving mixed states. Conventional quantum field theory techniques may not be fully appropriate in this context, necessitating the use of EFT methods tailored for finite-temperature systems.

\subsubsection{Effective Lagrangian for Quasicrystals}

We consider the effective action for quasicrystals in the classical limit, incorporating both phonon and phason degrees of freedom. The quasicrystal fields, denoted by $\psi^A$, represent internal rearrangements of the quasicrystal structure without displacing atoms. These fields are crucial as they can be interpreted as scalar fields in the EFT, making them ideal candidates for the inflaton field in cosmology.
The effective Lagrangian for quasicrystals, as presented in \cite{Baggioli:2020haa}, is given by:
\begin{equation} \mathcal{L}{\text{EFT}} = T^{\mu\nu} \partial\mu X_{a\nu} + J^{A\mu} \partial_\mu \psi_a^A + \Gamma^A \psi_a^A + \frac{i}{2} M^{AB} \psi_a^A \psi_a^B. \label{eqEFTLagrangian} 
\end{equation}
Here:
\begin{itemize} 
\item $T^{\mu\nu}$ is the stress-energy tensor, capturing the energy and momentum densities of the system. \item $J^{A\mu}$ are the currents associated with the quasicrystal fields, including both phonons and phasons. 
\item $\Gamma^A$ and $M^{AB}$ are dissipative coefficients introducing non-equilibrium dynamics and dissipation into the system. 
\item $X_{a\nu}$ and $\psi_a^A$ are the difference fields in the Schwinger-Keldysh formalism, defined as $X_{a\nu} = X_{1\nu} - X_{2\nu}$ and $\psi_a^A = \psi_1^A - \psi_2^A$, where the subscripts $1$ and $2$ denote the forward and backward branches of the contour, respectively. 
\end{itemize}

The field $\psi^4$ will be interpreted as the phason field. It transforms under a shift symmetry $\psi^4 \rightarrow \psi^4 + \lambda^4$, where $\lambda^4$ is a constant vector in the higher-dimensional space characterizing the quasicrystal. This shift symmetry implies that the phonon field behave as scalar fields under translations in the quasicrystal's higher-dimensional embedding space. Consequently, the phason field $\psi^4$ can be treated as scalar fields in the EFT framework, analogous to the inflaton field in cosmology.

\subsubsection{Phason Field as the Inflaton}

In our framework, the phason field $\psi^4$ acts as the inflaton field driving cosmic inflation. The choice of the component $\psi^4$ is motivated by its distinct role in the derivative expansion of the EFT, where its derivatives are counted differently compared to the spatial components $\psi^i$ ($i = 1, 2, 3$).

The dynamics of $\psi^4$ are governed by the equations of motion derived from the effective Lagrangian \eqref{eqEFTLagrangian}. Focusing on the phason component, the equation of motion for the classical (or retarded) phason field $\psi_r^4 = \frac{1}{2} (\psi_1^4 + \psi_2^4)$ is:
\begin{equation} H^{44} \partial_\mu \partial^\mu \psi_r^4 + \frac{\partial F^4}{\partial Z^4} \partial_0^2 \psi_r^4 - M^{44} \partial_0 \psi_r^4 = 0, \label{eqmotion} \end{equation}
where:
\begin{itemize} \item $H^{44}$ and $F^4$ are functions determined by the effective potential and kinetic terms in the Lagrangian. \item $Z^4 = \beta^\mu \partial_\mu \psi_r^4$, with $\beta^\mu$ being the inverse temperature four-vector. \item $M^{44}$ is a dissipative coefficient associated with the phason field. \end{itemize}

The equation of motion \eqref{eqmotion} describes the dynamics of a scalar field with dissipative terms, analogous to the inflaton field in cosmology. The presence of the dissipative coefficient $M^{44}$ introduces friction-like effects, which are significant in the context of warm inflation scenarios where the inflaton dissipates energy into radiation during inflation.

Transforming to Fourier space, we obtain the dispersion relation for the phason field:
\begin{equation} \omega^2 + i \gamma \omega = v^2 k^2, \label{eqdispersonphasonsomega} \end{equation}
where:
\begin{equation} \gamma = \frac{M^{44}}{H^{44} - \frac{\partial F^4}{\partial Z^4}}, \quad v^2 = \frac{H^{44}}{H^{44} - \frac{\partial F^4}{\partial Z^4}}. \end{equation}

The dispersion relation \eqref{eqdispersonphasonsomega} indicates that the phason field exhibits both propagating (real frequency) and dissipative (imaginary frequency) behavior. Using the relation $\omega = \dfrac{E}{\hbar}$ for the phason field, we rewrite the dispersion relation as:
\begin{equation} v^2 k^2 = \dfrac{E^2}{\hbar^2} + i \gamma \dfrac{E}{\hbar}. \label{eq:phasonrelation} 
\end{equation}
Here, $\gamma$ acts as the phason friction coefficient, representing the dissipative effects in the phason dynamics. This form explicitly shows how the energy $E$ of the phason relates to its momentum $k$, incorporating both the dispersive term $\dfrac{E^2}{\hbar^2}$ and the dissipative term $i \gamma \dfrac{E}{\hbar}$.

The phason field's behavior as a scalar field with dissipative dynamics makes it an attractive candidate for the inflaton in cosmology. The shift symmetry of $\psi^4$ ensures that it can have a flat potential, a key requirement for slow-roll inflation. Additionally, the dissipative term allows the phason field to decay and transfer energy to matter fields during inflation, leading to particle production and reheating within the inflationary period.

\subsubsection{Built-in Reheating Mechanism}

The imaginary term $\frac{i}{2} M^{AB} \psi_a^A \psi_a^B$ in the Lagrangian introduces dissipation, which leads to particle production during the inflationary period. As the phason field evolves, it decays into matter and radiation fields through typical couplings present in the quasicrystal EFT. This process effectively reheats the universe without the need for a separate reheating phase, providing an alternative to standard reheating scenarios.

The decay rate of the phason field can be inferred from the dissipative coefficients $M^{AB}$ in the Lagrangian. The reheating temperature $T_{\text{reh}}$ can be estimated by considering the energy density transferred to radiation during the phason's decay. This mechanism seamlessly transitions the universe from the inflationary phase to the radiation-dominated era, aligning with observational constraints on the reheating temperature.

\subsubsection{Implications for Cosmic Inflation}

By utilizing the phason field inherent in quasicrystals, we propose a novel approach to cosmic inflation that naturally incorporates dissipation and particle production. The unique properties of the phason field offer several advantages:

\begin{itemize} \item \textbf{Scalar Field Dynamics}: The phason field $\psi^4$ behaves as a scalar field in the EFT, with its dynamics governed by derivative interactions and shift symmetry. This aligns with the requirements for the inflaton field in cosmology. \item \textbf{Slow-Roll Inflation}: The shift symmetry of the phason field ensures a flat effective potential, satisfying the slow-roll conditions necessary for a prolonged period of accelerated expansion. \item \textbf{Intrinsic Dissipation and Reheating}: The dissipative terms in the effective Lagrangian lead to friction-like effects, causing the phason field to lose energy to particle production during inflation. This provides a built-in reheating mechanism that naturally transitions the universe to the hot, radiation-dominated phase. \item \textbf{Compatibility with Observations}: The model can be tuned to produce a spectrum of primordial perturbations consistent with cosmic microwave background observations. The dissipative dynamics may also lead to unique signatures in the non-Gaussianities of the perturbations, offering potential observational tests of the model, a work in progress. \end{itemize}
---

To review our working hypothesis that the universe is a growing quasicrystal, we propose an early quantum mechanical phase,  
followed by the natural emergence of phonons and phasons. Once the quasicrystal grows sufficiently large, such that the dynamics occur on distance scales much larger than the spacing between the underlying points, we expect the model to exhibit continuous symmetry in the tiling space. At this stage, effective field theory can be applied, with phonons and phasons modeled as scalar fields. The interplay between phonon and phason modes may contribute to the generation of mass scales through mechanisms akin to dynamical symmetry breaking, where collective excitations lead to effective masses for particles.

\subsection{Phonons and Dark Matter}

In our quasicrystalline universe model, we extend the analogy with solid-state physics to propose a novel explanation for dark matter. The phonon fields $\psi^i$ in Equation (\ref{eqEFTLagrangian}) correspond to collective excitations (vibrations) of the quasicrystal lattice in spacetime. Unlike phasons, these fields follow conventional dispersion relations and could survive as relics, interacting weakly with ordinary matter and potentially contributing to the dark matter component of the universe.

Much like axions \cite{Preskill, Abbott, Baggioli2021}, which are not thermally produced but arise as classical oscillations of the axion field, giving rise to cold dark matter, phonons in our model emerge from the quasicrystal field through an EFT framework. Both axions and phonons are characterized by long wavelengths, enabling them to contribute to the dark matter component of the universe while remaining difficult to detect through electromagnetic interactions.

Additionally, just as WIMPs are theorized to interact weakly with ordinary matter, phonons in the spacetime quasicrystal could potentially scatter off nuclei, producing recoil signals detectable in experiments designed to search for WIMPs. The weak interaction between spacetime phonons and ordinary matter presents a similar observational challenge to that of axions and WIMPs, offering an alternative mechanism for dark matter.

By drawing on these analogies, we propose that spacetime phonons could serve as dark matter candidates. They would influence the dynamics of ordinary matter while remaining largely undetectable, providing a novel perspective grounded in the quasicrystalline structure of spacetime proposed by our model.

\subsection{Growth Dynamics in a Quasicrystalline Universe and Dark Energy}

Our quasicrystalline universe model incorporates two key growth mechanisms:

\paragraph{External Growth by Shell Projection}

The universe expands externally through the projection of new points from a higher-dimensional lattice onto our observable space. Specifically, we consider the projection of the 8-dimensional $E_8$ lattice onto 4-dimensional space to form the ESQC. This process adds concentric shells of points, effectively growing the quasicrystal—and thus the universe—shell by shell.

\paragraph{Internal Growth through Subdivision and Rescaling}

As new points are projected, they may appear within existing structures, leading to internal subdivision. Due to the quasicrystal's self-similarity, the tile structures are preserved at different scales. To maintain consistency with a minimal length scale imposed by matter (e.g., the Planck scale), the universe undergoes \emph{rescaling}. Whenever new points introduce distances smaller than this minimal scale, a global rescaling adjusts all structures to preserve fundamental physical constants.

This mechanism can be more easily grasped in lower-dimensional quasicrystals, such as the Fibonacci chain (one-dimensional) and the Penrose tiling (two-dimensional), which can be built with just two tile types. Steinhardt's one-tile quasicrystal~\cite{Steinhardt1996} further demonstrates how tiles keep dividing but reappear at different scales due to self-similarity. In three dimensions, quasicrystals like the 3D Penrose tiling involve more complex arrangements but follow the same principle~\cite{Duneau1985}. In four dimensions, with the ESQC, the number of tiles is not well known, but the building block is the 600-cell polytope, which keeps arising at different scales.

Whenever a new shell introduces a shorter minimum distance, a global rescaling is applied to the entire quasicrystal. This rescaling adjusts all distances to ensure consistency with the previously established minimum distance, preventing distances from shrinking below a critical threshold and ensuring that the quasicrystal maintains a fixed internal structure.

\begin{figure}[ht]
    \centering
    \includegraphics[scale=0.40]{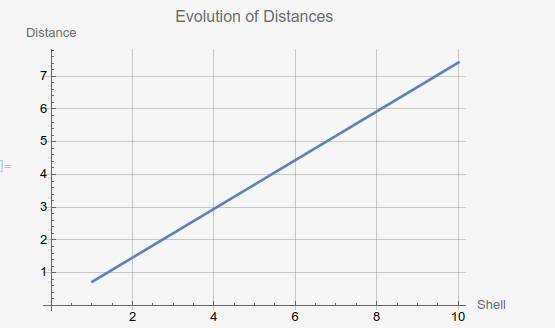}
    \caption{The image demonstrates the stepwise increase in distance between two points in the ESQC, with rescaling applied when necessary. We project the first 10 shells of the E8 lattice one by one into 4D space. The distance from the origin in the first shell is fixed as the minimum reference distance. As subsequent shells are projected, new points may introduce shorter distances, and whenever this occurs, a global rescaling is applied to maintain the initial minimum distance. The linear progression in the image reflects the cumulative effect of this rescaling mechanism as the quasicrystal grows.}
    \label{e8to4drescaling}
\end{figure}

\paragraph{Implications for Dark Energy}

The rescaling process requires energy to adjust the quasicrystalline structure while maintaining consistency with physical laws consistent. This energy can be associated with an effective \emph{dark energy} component, contributing to the cosmological constant $\Lambda$ in the Friedmann equations. The continuous introduction of new points and the accompanying rescaling provide a natural mechanism for dark energy arising from the quasicrystalline structure of spacetime.

\paragraph{Addressing the Hubble Tension}

The dual growth mechanisms may impact cosmological observations differently over time:

\begin{itemize} \item \textbf{Early Universe}: External growth dominates, with rapid shell projections expanding the universe. \item \textbf{Later Universe}: Internal growth and rescaling become significant as matter influences the minimal scale, potentially affecting the observed expansion rate. \end{itemize}

This interplay offers a qualitative framework to address the \emph{Hubble tension}, the discrepancy between early and late-time measurements of the Hubble constant, by suggesting that variations in the expansion rate are tied to the quasicrystalline growth dynamics.

\section{From Schr{\"o}dinger to Friedmann Equation}
\label{sec:freedmaneqwavefunction}

In the previous sections, we proposed a heuristic model where the cosmic expansion observed at large scales emerges intrinsically from the growth dynamics of a quasicrystalline structure underlying spacetime. We considered how this quasicrystalline universe grows both externally, through the projection of new points from higher dimensions, and internally, via subdivision and rescaling to maintain consistency with a minimal length scale. Now, we aim to address how this theoretical growth, even at small scales, might affect quantum matter.

General relativity establishes a correspondence between the geometric aspects of spacetime and the distribution of matter and energy, as encapsulated by Einstein's local field equations,
\begin{equation}
G_{\mu\nu} + \Lambda g_{\mu\nu} = \frac{8\pi G}{c^4} T_{\mu\nu},
\end{equation}
where $G_{\mu\nu}$ is the Einstein tensor representing the curvature of spacetime, $g_{\mu\nu}$ is the metric tensor describing the geometry of spacetime, $T_{\mu\nu}$ is the stress–energy tensor reflecting the distribution and flow of energy and momentum in spacetime, $\Lambda$ is the cosmological constant, $G$ is the Newtonian constant of gravitation, and $c$ is the speed of light in vacuum. 

On the other hand, quantum mechanics introduces the principle of wave-particle duality, which manifests both wave-like and particle-like properties in matter and energy, as illustrated by the De Broglie relations and the Schr{\"o}dinger equation-Equations (\ref{eq1Schrodinger}, \ref{eq2KEV}). 

Notably, both General Relativity and Quantum Mechanics relate matter to an underlying structure—geometric in the case of spacetime curvature, and wave-like in the case of quantum wavefunctions. This observation motivates us to explore a simplified model within the cosmological context, where the wave aspect of quantum mechanics is proportionally linked to the geometry of spacetime.

The Friedmann equations, which are key to understanding the universe's expansion, are derived from general relativity under the assumption of a homogeneous and isotropic universe. These equations describe how the scale factor a(t) evolves over time due to the energy content of the universe. Our aim is to investigate whether analogous dynamics could emerge from the effects of spacetime rescaling on local quantum states, focusing on the later evolution in flat space at small quantum scales. The early universe's curvature's regime will be addressed in a separate paper.

Conventionally, the expansion of the universe is significant at large scales, particularly in the regions between galaxies, while gravitational forces within galaxies and smaller structures prevent any noticeable effect. However, this does not rule out the possibility that spacetime expansion occurs even within matter-dominated regions, albeit in ways that are difficult to observe directly. To explore this, we examine how a quantum state behaves in an expanding universe.

Here we consider the flat space scenario. The time-independent Schr{\"o}dinger Equation for the wave function (energy eigenfunction) $\Psi(x)$, with $x$ representing $(x,y,z)$, of a particle with mass $m$ under potential $V_{QM}(x)$. The equation is given by:
\begin{equation}
\nabla^2 \Psi = -\frac{2m(E - V_{QM})}{\hbar^2} \Psi \quad, \label{eq1Schrodinger}
\end{equation}
with $E$, the energy eigenvalue for the stationary state $\Psi(x)=\Psi_{E}(x)$.

For a particle under a potential $V_{QM}(x)$, the time-independent Schr{\"o}dinger equation can yield different forms of solutions depending on the relationship between the energy $E$ of the particle and the potential $V_{QM}$. We consider the particle in a box situation with a finite potential inside and infinite outside. The box may be taken to be spherical with a given radius. Specifically, inside the box, for a constant potential, being zero, or when $E > V_{QM}$, the solutions are of the form of plane waves. Conversely, for $E < V_{QM}$, the solutions differ significantly. When these solutions are substituted back into the time-independent Schr{\"o}dinger equation, we obtain:
\begin{equation}
k^2 = \frac{2m(E-V_{QM})}{\hbar^2}, \label{eq2KEV}
\end{equation}
where $k$ is the wavenumber.

This is a stationary state analysis. We now consider a scenario where spacetime itself begins to change, specifically, rescaling with a factor $a(t)$. Consequently, the volume of a given region $\eta(t_1)$ at time $t_1$ will relate to the volume $\eta(t_2)$ at time $t_2$ by
\begin{equation}
\eta(t_2) = \frac{a(t_2)^3}{a(t_1)^3}\eta(t_1), \label{eq:VolumeScale}
\end{equation}
ensuring that the ratio $\eta_0 = \frac{\eta(t)}{a(t)^3}$ remains constant. Thus, we can express $\eta(t)$ as $\eta_0 a(t)^3$.

Introducing time is essential to account for the rescaling effects. The problem now should be addressed by the time-dependent Schr{\"o}dinger equation, as the potential depends on the dynamic radius or volume. This problem has been studied using various approaches \cite{Cooney} and references therein. In this context, the energies become dynamic, and a notion of energy density in quantum mechanics arises naturally \cite{francisco}. Crucially, the wavenumber, wavelength, and quantum amplitudes become functions of the dynamical size of the box, and we posit that they are subject to the Hubble-Lemaître's law explicitly. By connecting the geometry with the wave aspect of the particle-wave duality mentioned earlier, we aim to derive Friedmann-like equations. While Hubble-Lemaître's law is empirically validated at cosmological distances, we extend its conceptual framework to 'intermediate' or even microscopic scales as a theoretical proposition. These scales are not conventionally associated with cosmic expansion, but within our growing quasicrystal model, the underlying spacetime structure allows for such an extension. This hypothesis serves as a foundation to explore the consequences of spacetime expansion on quantum states confined within matter-dominated regions.

Revisiting equation (\ref{eq2KEV}), we modify it by dividing both sides by the dynamical volume $\eta$, which confines the particle in space. Equation (\ref{eq2KEV}) encapsulates the wave-particle duality, highlighting that the matter wavelength, $\lambda = \frac{2 \pi}{k}$, is a fundamental quantum mechanical attribute that determines the probability density of locating the particle within the configuration space. Notably, a particle's wavelength, inversely related to its momentum (and thus energy), serves as a bridge to cosmological dynamics. By postulating the Hubble-Lemaître's law at small scales, expressed as $v = H \lambda$, where $H = \frac{\dot{a}}{a}$ is the time-dependent Hubble parameter, we correlate the particle's wavelength to the cosmic scale factor $a(t)$ while keeping the energy constant. 
In our analysis, $\lambda$ represents the instantaneous wavelength of the particle's wavefunction within the expanding box. It is the `local' wavelength affected by the expansion of the confinement volume, rather than an observed wavelength subject to cosmological redshift. Since we are modeling the particle within a dynamically expanding potential, $\lambda$ changes as a function of time due to the expansion, reflecting how the particle's quantum state adapts to the changing spatial boundaries. We focus on this local effect rather than the observed redshift  because we are interested in how the expansion influences the quantum properties of the particle itself, not how an external observer perceives changes in wavelength over cosmological distances.
This association suggests an emergent \emph{effective kinetic energy} tied to the velocity vv associated with the expansion. We acknowledge that, in standard cosmology, recession velocities resulting from the expansion of space are not associated with kinetic energy in the Newtonian sense, as they do not correspond to motion through space but rather to the metric expansion of space itself. As discussed by Sharma \cite{Sharma2024}, kinetic energy calculations involving galaxies must consider both peculiar velocities (which contribute to kinetic energy) and recessional velocities due to cosmic expansion (which do not contribute to kinetic energy in the traditional sense).
However, within our theoretical framework, we consider the impact of the expansion on a localized quantum system confined within an expanding potential. Here, we introduce the \emph{effective kinetic energy} as a conceptual parameter tied to the expanding volume, representing the energy associated with the expansion-induced changes in the quantum system's boundary conditions. This effective kinetic energy is not due to the particle moving through space in the classical sense but emerges from the changing spatial confinement resulting from the expansion of the universe.

Consequently, equation (\ref{eq2KEV}) is transformed into
\begin{equation}
H^2 = \left( \frac{\dot{a}}{a} \right)^2 = \frac{K\eta}{\pi^2 \hbar^2} \rho, \label{eq:Friedmann1fromQuantum}
\end{equation}
where $K=\frac{1}{2}mv^{2}$, $\rho=E/\eta$ represents the energy density, and setting $V_{QM}=0$. Note that we are simplifying the concept of energy density here. The particle energy is time-dependent, and any particle would be subject to the same expansion that occurs everywhere. This brings a notion of a particle horizon and an energy density within each particle volume expansion. This equation bears a striking resemblance to the first Friedmann equation, suggesting that the dynamics of a quantum particle in a box with time-dependent boundaries may provide insights into the expansion of the universe by establishing a connection between the particle's wave nature and the cosmic scale factor. By using the effective kinetic energy as a conceptual tool, we aim to explore the potential interplay between quantum mechanics and cosmology within our model, acknowledging that this is a theoretical extension and not empirically verified at small scales.
To align with the usual first Friedmann equation's structure, the term $\frac{K\eta}{\pi^2 \hbar^2}$ must remain constant
\begin{equation}
K\eta=\frac{8\pi^{3}\hbar^{2}G}{3c^{2}} \label{kineticvolumeconstar}
\end{equation}
with kinetic energy compensating for the changing volume and establishing a direct quantum mechanical analogy to cosmic expansion dynamics.
\begin{equation}
H^2 = \left( \frac{\dot{a}}{a} \right)^2 = \frac{8\pi G}{3c^{2}} \rho, \label{eq:Friedmann1fromQuantum2}
\end{equation}

The analysis of a relativistic particle in a box with moving walls is similar to, but more complex than, the non-relativistic case. This complexity arises whether we consider a fermionic spin-1/2 particle described by the Dirac equation or a spin-0 particle described by the Klein-Gordon equation \cite{Koehn:2012red}. To gain insights into the relativistic case, we can apply the same analysis using the relativistic dispersion relation derived from the energy-momentum relation:
\begin{equation}
(\hbar kc)^2  = E^2 - (mc^2)^2. \label{eqRelDispersion}
\end{equation}

To bridge the quantum and classical realms, we need to determine the energy density. This involves multiplying and dividing the equation by the dynamical volume $\eta$. In the non-relativistic case, we notice the appearance of a kinetic energy term linked to the expansion. This observation suggests that the dynamical energy $E$, for a unique particle, can transform into different forms during the expansion process. Specifically, in the case of photons, where, $mc^2=0$, we consider the energy density related to the photon's energy, which evolves with the expansion. Motivated by this observation, we apply the energy-balancing constraint:
\begin{equation}
\ensuremath{C=E-K-V=0},\label{eq:energybalanceconstraint}
\end{equation}
where $E$ is the total energy, $K$ is a effective kinetic energy, and $V$ is a effective potential energy, all associated with the dynamical expansion process. This constraint allows us to examine the system's properties across different scales, connecting the microscopic energy $E$ to macroscopic scales potentially represented by $K$ and $V$. Here, $K$ and $V$ can be viewed as effective parameters encapsulating average or microscopic system properties as the system expands.

The energy-balancing constraint in Equation (\ref{eq:energybalanceconstraint}) provides a framework for understanding how the total energy $E$ can be viewed in the expansion side as partitioned into kinetic and potential components during the expansion process. By considering the conservation of energy and the interplay between these different forms of energy, we can gain insights into the dynamics of the system across various scales.
In the context of the relativistic particle in a box with moving walls, the application of this constraint may lead to the emergence of terms analogous to the energy density $\rho$, pressure $p$, and possibly the cosmological constant $\Lambda$ when deriving the relativistic Friedmann-like equation. By incorporating the energy-balancing constraint into the analysis, we can explore how the quantum mechanical properties of the system, such as the wave nature of the particle and the dynamical evolution of the wavefunction, relate to the classical cosmological parameters that describe the expansion of the universe.

Isolating $K$ while keeping $E$ and $V$ independent, we obtain 
\begin{equation}
E^{2}=K(E+V+V^{2}/K).\label{E2expanded}
\end{equation}
Since the wave number is inversely proportional to the wavelength, we can apply the Hubble-Lemaître's law as done before.  Then, Equation (\ref{eqRelDispersion}) can be rewritten as:
\begin{equation}
H^{2}=\beta^{2}\frac{K\eta}{\pi^{2}\hbar^{2}}(\rho+p)+\frac{\beta^{2}}{\pi^{2}\hbar^{2}}V^{2}-\frac{\beta^{2}}{\pi^{2}\hbar^{2}}(mc^{2})^{2}\label{eqrelativisticH}
\end{equation}
with $\beta=v/c$, $\rho = E/\eta$ is the energy density, and $p = V/\eta$ is the pressure. Equation (\ref{eqrelativisticH}) is analog of the Friedmann equation for relativistic matter, derived from the quantum mechanical description of a particle in a box with moving walls. The first term on the right-hand side resembles the classical Friedmann equation. The second term, proportional to $V^2$, may be interpreted as related to the cosmological constant $\Lambda$ as it does not explicitly depend on $a$.

To obtain the usual contribution to radiation, we set $mc^2 = 0$. Typically, radiation energy density enters the Friedmann equation as proportional to $1/a^4$, while non-relativistic matter is proportional to $1/a^3$. Here, we derive a notion of energy density for individual massive particles and photons subject to dynamical boundary conditions related to the expansion of space, and they share a similar form. The $1/a^4$ factor for radiation arises from considering the number density of photons, which scales like matter, times the photon energy, which scales with $1/a$ due to redshift as the photon wavelength expands. The redshift here is accounted for through the Hubble-Lemaître's law applied to the wavelength.

The essential difference between non-relativistic matter and radiation is that in one case $H^2 \propto E$, and in the other, $H^2 \propto E^2$. To achieve the same notion of energy density, our approach introduces a new term for radiation. Essentially, we achieve the same result of having different phases: one in the short wavelength regime and the other at large wavelengths at later times.

Importantly, we note that different dispersion relations will lead to different dynamics for the scale parameter, and various relations may be considered. This is in line with the usual approach where different cosmological epochs have different evolutions due to varying energy densities.

Let us visit now the dispersion relation for phonons, Equation (\ref{eq:phasonrelation}). In the case of lattices, the energy states are defined in a precise polytope of the lattice, the Voronoi cell. For quasicrystals, there are more fundamental cells or tiles, but with one fixed, we can consider the expansion of that cell and, as before, apply the Hubble-Lemaître's law to the wavelength. We also need to take in account the different regimes of short and large wavelength. 
To deal with the imaginary part, this equation can be interpreted from the energy relation where in the regime of short wavelength the phason propagation dominates with $v^{2}k^{2}=\frac{E^{2}}{\hbar^{2}}$ leading to a Friedmann like equation similar to Equation (\ref{eqrelativisticH}) as we can use Equation (\ref{E2expanded}) and multiply by the volume on both sides of Equation (\ref{eq:phasonrelation}), yielding 
\begin{equation}
H^{2}=\frac{K\eta}{4\pi^{2}\hbar^{2}}(\rho+p)+\frac{1}{4\pi^{2}\hbar^{2}}V^{2} \label{phasonH2}
\end{equation}
where we considered the velocity associated to phasons equal to the one of the expansion. 

In the regime of large wavelength we have $v^{2}k^{2}=i\gamma\frac{E}{\hbar}$ which leads to diffusion and decoupling from expansion contribution
\begin{equation}
H^{2}=i\gamma\frac{E}{4\pi^{2}\hbar} \label{phasonHimaginary}
\end{equation}

To finish this analysis, we can add the acceleration equation from usual thermodynamics considerations when one derives the Friedmann equations from Newtonian physics. For this, one adds the continuity equation
\begin{equation}
\dot{\rho} + 3 \frac{\dot{a}}{a} (\rho + p) = 0.
\end{equation}
Taking the derivative in Equations (\ref{eq:Friedmann1fromQuantum}), (\ref{eqrelativisticH}), and (\ref{phasonH2}), with $V$ constant and $\beta=1$, we get for Equations (\ref{eq:Friedmann1fromQuantum}), (\ref{eqrelativisticH}) and (\ref{phasonH2}):
\begin{equation}
2H \dot{H} = \alpha \dot{\rho},
\end{equation}
with $\alpha = K\eta/\pi^2 \hbar^2$. 
Now, we can derive the acceleration term in both situations. For Equation (\ref{eq:Friedmann1fromQuantum}), substituting the continuity equation into the derivative of the Friedmann-like equation yields (with the original potential equal to zero)
\begin{equation}
\frac{\ddot{a}}{a} = -\frac{\alpha}{2} (\rho + 3p).
\end{equation}
This equation describes the acceleration of the scale factor in terms of the energy density and pressure, analogous to the second Friedmann equation in general relativity.

For Equations (\ref{eqrelativisticH}), with $m=0$
\begin{equation}
\frac{\ddot{a}}{a}=-\frac{\alpha}{2}(\rho+p)+\Lambda \label{freedmanrela1}
\end{equation}
with $\Lambda=\frac{1}{\pi^{2}\hbar^{2}}V^{2}$.

For Equation (\ref{phasonH2}), the acceleration equation becomes
\begin{equation}
\frac{\ddot{a}}{a}=-\frac{\alpha}{4}(\rho+p)+\frac{1}{4}\Lambda \label{freedmanrela2}
\end{equation}

These acceleration equations provide insights into the dynamics of the scale factor in the presence of matter, radiation, and phason contributions. They demonstrate how the quantum mechanical description of particles in a box with moving walls can give rise to cosmological equations that resemble the Friedmann equations of general relativity. The inclusion of the phason term highlights the potential role of quasicrystalline structures and their associated quasiparticles in the evolution of the universe.

Our analysis commenced with the time-independent Schr{\"o}dinger equation for a particle confined within a potential well. We then introduced several crucial modifications to the system: a general energy-balancing constraint Equation (\ref{eq:energybalanceconstraint}), a time-dependent boundary condition representing moving walls, and a dynamically evolving volume. These additions were aimed at establishing a connection between the quantum and classical regimes across different scales. By applying the Hubble-Lemaître's law to the wavelength of the particle and relating it to the cosmic scale factor, we derived a Friedmann-like equation Equation (\ref{eq:Friedmann1fromQuantum}) from the quantum mechanical description of the system.

In the relativistic extension of our analysis, we employed the Klein-Gordon equation and the relativistic energy-momentum relation Equation (\ref{eqRelDispersion}). This led to the emergence of a new term in the resulting Friedmann-like equation Equation (\ref{eqrelativisticH}), which bears a resemblance to the cosmological constant term in the standard Friedmann equations of general relativity.

Furthermore, we explored the implications of different dispersion relations, particularly those associated with phonons and phasons in quasicrystalline structures. The presence of an imaginary term in the phason dispersion relation Equation (\ref{eq:phasonrelation}) is interpreted as having different regimes contributions. 

In summary, our analysis bridges the quasicrystal model of the universe with quantum mechanical systems, showing how the underlying expanding quasicrystalline structure affects the behavior of quantum matter. We demonstrated that, within this framework, there exists a duality between quantum states, seen either as wave-like matter or as an expression of the evolving geometry through the scale parameter. This duality draws a parallel with Einstein's vision, where matter and geometry are deeply interconnected. By relating the matter wavelength to the expansion of space, we linked the Schr{\"o}dinger equation to the Friedmann equation. This suggests that the classical world’s emergence may be intrinsically tied to the expanding substrate, where we perceive either matter or expansion, but not both simultaneously at local scales. The quantum mechanical description naturally adapts to this evolving background, providing new insights into the relationship between quantum states and cosmic expansion.  We also note that the phonon contribution in the early short-wavelength regime introduces a term, $1/4\Lambda$, which may offer a potential resolution to the Hubble tension, another workin in progress.

\section{Discussions, Exploratory Insights and Future Perspectives}
\label{sec:conclusion}

\subsection{Implications of the Universe as a Growing Quasicrystal}

The hypothesis that the expanding universe is a growing quasicrystal has several significant implications for our understanding of cosmology:

\begin{enumerate} \item \textbf{Alternative to Inflation:} The need for an inflationary period may be reconsidered. In our model, the large number of tiles or points $N$ required for a uniform distribution is achieved inherently through the quasicrystalline structure. This intrinsic uniformity of quasicrystals could provide a natural explanation for the observed large-scale homogeneity and isotropy of the universe without invoking an inflationary phase.

\item \textbf{Phasons as Scalar Fields and Reheating:} Following the initial quantum growth phase, the universe transitions to an epoch characterized by the emergence of phasons and phonons—quasiparticles associated with the quasicrystalline structure. Phasons, in particular, can be modeled as scalar fields similar to the inflaton in conventional cosmology. Their dynamics can contribute to reheating processes, generating matter and radiation in the early universe. The unique dispersion relations of these quasiparticles influence the energy density and expansion rate during this period.

\item \textbf{Diminishing Phason Influence at Later Times:} As the universe continues to grow and the number of points \( N \) increases, the contribution of phasons to the energy density becomes negligible. This is due to the dominance of a diffusive imaginary term in their dispersion relation at large wavelengths, effectively damping their influence on the universe's expansion in later epochs.

\item \textbf{Role of Matter and Long-Wavelength Phonons:} Matter, along with long-wavelength massive phonons, remains critical in the universe's evolution. These components contribute significantly to the overall energy density and can interact with ordinary matter. Long-wavelength massive phonons, in particular, could act as candidates for dark matter, influencing structure formation and cosmic dynamics.

\item \textbf{Accelerated Expansion and Internal Rescaling:} The universe's accelerated expansion can be explained by an internal rescaling mechanism driven by a fixed matter scale. This rescaling results from the continuous projection and replacement of points within the quasicrystalline structure. It effectively acts as a form of dark energy or cosmological constant, driving the observed acceleration in the universe's expansion without requiring additional exotic energy components.

\item \textbf{Emergent Symmetries from Discrete Structures:} While exact invariant states in a discrete spacetime setting may be challenging to identify \cite{bojowald}, our quasicrystalline model suggests that large-scale symmetries emerge as approximations. By applying a statistical mechanics framework to the tiling space microstates, we can derive partition functions and energy densities. These lead to Friedmann-like equations when we impose the Hubble-Lemaître's law on the wavelength in the dispersion relations of the propagating degrees of freedom. Thus, continuous symmetries can emerge from the underlying discrete quasicrystalline structure at macroscopic scales.
\end{enumerate}

Overall, this model provides a novel framework for understanding the universe's structure and dynamics, integrating concepts from condensed matter physics and cosmology. It offers fresh perspectives on longstanding issues such as the Hubble tension and the nature of dark matter and dark energy.

\subsection{Testing the Quasicrystalline Universe}

At this juncture, readers may question why projections occur, how the system determines which shells to project, and the nature of the computational framework underlying this higher-dimensional lattice. They may also wonder if higher-dimensional lattices are involved beyond the initial construction. These inquiries parallel conventional physics questions about the Big Bang, conservation laws, and the fundamental mechanisms underlying physical laws. Despite these profound questions, the hypothesis of a higher-dimensional framework introduces testable consequences for the cosmos—namely, the emergence of a quasicrystalline structure distinct from both continuous and traditional crystalline constructs.

If the cosmos possesses a quasicrystalline structure, this feature could potentially be identified using techniques similar to those employed in materials science for detecting crystalline arrangements. The structural nature of a material—whether crystalline, quasicrystalline, or amorphous—is revealed through distinctive diffraction patterns. In the laboratory, such patterns are commonly produced through electron diffraction, where an electron beam interacts with the material. The resultant diffraction pattern provides detailed information on atomic spacing and structure. The presence of definitive Bragg peaks within these patterns indicates constructive interference from regular atomic planes, revealing the material's symmetry and structure.

Applying this concept on a cosmic scale, one might consider the CMB radiation as a macroscopic ``diffraction pattern'' reflecting the universe's structure. The CMB, a relic radiation from the early universe, offers a primordial snapshot from approximately 380,000 years after the Big Bang, containing valuable structural and compositional data. Detailed examinations of temperature and polarization fluctuations within the CMB, as captured by observatories like the Planck satellite, allow researchers to infer the universe's large-scale structure.

In a theoretical model positing quasicrystalline symmetry in the universe, this analysis would involve searching the CMB data for Bragg-like peaks—specific intensity patterns revealing an ordered structure characteristic of quasicrystals. Identifying such patterns would have profound implications, potentially transforming our understanding of the cosmos and bridging celestial phenomena with quantum-scale observations in condensed matter physics.

\subsection{Quantum Error-Correcting Codes}
The growing Hilbert space structure presented earlier suggests that the early universe may be viewed as a form of a quasicrystalline topological quantum computer, with quantum error-correcting codes (QECCs) inherently embedded \cite{Li:2023nkr}. At each growth step, a subspace $\mathcal{C}_n \subset \mathcal{H}_n$ can be identified, spanned by a set of wavefunctions ${|\Psi{n,[T]}\rangle}$ that exhibit local indistinguishability and recoverability properties \cite{Li:2023nkr}:
\begin{equation}
|\Psi_{n,[T]}\rangle = \sum_{g \in G_n} \beta(g T_n) |g T_n\rangle,
\end{equation}
where $[T]$ denotes an equivalence class of tiling configurations related by the symmetry group $G_n$, and $\beta(g T_n)$ are complex coefficients. The subspace $\mathcal{C}_n = \text{span}({|\Psi_{n,[T]}\rangle})$ can be interpreted as the code space of a QECC at the $n$-th growth step, capable of correcting errors in any local region of the quasicrystal. As the universe grows, the QECC evolves, $\mathcal{C}_n \rightarrow \mathcal{C}_{n+1}$.

This growing QECC structure provides a framework for understanding how quantum information can be encoded and protected within the geometry of a pre-inflation quasicrystalline universe. Notably, the Hilbert spaces associated with certain quasicrystals are isomorphic to the fusion Hilbert spaces that appear in topological quantum computing systems \cite{Amaral2022sy}. This establishes a connection between QECCs and topological quantum computation within the context of our working hypothesis of the growing quasicrystalline universe.

%The emergence of this quasicrystalline structure and its associated Hilbert spaces in the early universe may have profound implications for the evolution of spacetime, matter, and quantum information processing. As the universe continues to grow and evolve, the interplay between the quasicrystalline geometry, QECCs, and topological quantum computation may guide the formation of large-scale cosmic structures and the emergence of complex phenomena. This heuristic model offers a novel perspective on the origin and evolution of the universe, connecting ideas from condensed matter physics, quantum information theory, and cosmology.

\subsection{Conclusion}

In this paper, we have proposed a novel heuristic model of the expanding universe conceptualized as a quasicrystalline projection from a higher-dimensional lattice. By presenting the growing quasicrystal model first, we have established a foundational framework where the intrinsic growth dynamics of a quasicrystal can naturally give rise to the observed cosmic expansion. The inherent properties of quasicrystals, which resonate with cosmological principles such as self-similarity and aperiodic order, offer promising insights into the dynamics of the cosmic scale factor.

Our model envisions the universe's scale factor as a function of both the non-linear increase in points resulting from projections from a higher-dimensional lattice and the rescaling of spaces within the matter content of the universe as new points are added to the projected tiling space. This dual growth mechanism provides an early phase characterized by quantum error correction, potentially bypassing the need for an inflationary phase. The universe then transitions to a phase dominated by propagating phasons, phonons, and radiation, eventually evolving to a matter-dominated phase in the limit of a very large number of tiles. In later epochs, the universe experiences two kinds of expansion mechanisms due to the continuous projections and internal rescaling. Long-wavelength massive phonons emerge as candidates for dark matter, offering a fresh perspective on its nature and role in cosmic evolution.

Motivated by the intrinsic quasicrystal growth, we explored the Schr{\"o}dinger equation for a particle in both flat and curved spaces. By considering a time-dependent potential arising from the quasicrystal's expansion, we introduced a constraint that bridges microscale quantum phenomena with macroscale cosmological quantities. This approach led to the derivation of an equation bearing a structural resemblance to the Friedmann equation, which governs the dynamics of the universe's expansion in General Relativity.

This development could be instrumental in addressing pivotal cosmological challenges, such as the \emph{Hubble tension}. The Hubble tension arises due to observed discrepancies in the measurement of the universe's expansion rate, as denoted by the Hubble constant H0. Notably, H0 shows differing values when measured in the early universe—particularly via the CMB, indicating a slower expansion rate—compared to measurements from the late universe using supernovae and Cepheid variables, which suggest a faster expansion rate. Such variations point to potential gaps in our current understanding of the universe's expansion history. Our model offers a new avenue to reconcile these differences by providing a mechanism where the expansion rate naturally varies due to the intrinsic properties of quasicrystal growth.

A potential critique of this model could be the apparent contradiction between the continuous symmetries observed in the universe and the discrete nature of a quasicrystalline structure. However, this can be addressed by noting that quasicrystals give rise to tiling spaces that exhibit continuous symmetries on very large scales. We introduced a statistical mechanics framework to demonstrate how continuous behavior emerges from the underlying discrete structure. By applying a probabilistic measure to the tiling space microstates, we derived partition functions and energy densities. The Friedmann-like equations arose by imposing the Hubble-Lemaître's law on the wavelengths in the dispersion relations of propagating degrees of freedom, showing that large-scale cosmological dynamics can be consistent with a fundamentally discrete spacetime.

Furthermore, the exploration of the Schr{\"o}dinger equation within this framework underscores the interplay between quantum mechanics and cosmology. It suggests that quantum states are influenced by the expansion of spacetime inherent in the quasicrystal growth, potentially leading to observable consequences in the behavior of quantum particles. This connection between microscale and macroscale phenomena enriches our understanding of the universe and highlights the importance of considering novel models that integrate concepts from different areas of physics.

To fully realize the potential of this approach, further detailed analysis is required. Future work should focus on rigorously developing the quantitative aspects of the model, exploring its predictive power, and examining its compatibility with observational data. 
By advancing this research, we aim to contribute significantly to our understanding of the universe's expansion and its underlying structure. The integration of quasicrystal mathematics with quantum mechanics and cosmology offers a rich and promising framework that could shed light on some of the most profound mysteries in physics.

\end{document}